\documentclass[aps,prb,twocolumn,noshowpacs,superscriptaddress]{revtex4-2}
\usepackage{amsmath,graphicx,amsbsy,amssymb,amsmath,epsfig,latexsym,wasysym,pifont,mathrsfs,natbib}
\usepackage{float} 
\newfloat{widefig}{thp}{lop} 
\usepackage{comment}
\usepackage{soul}
\usepackage{array, makecell} \usepackage{verbatim} \usepackage{datetime}
\usepackage{multirow,array} \usepackage{hyperref} \usepackage{color} \usepackage{setspace}
\usepackage{graphicx}
\usepackage{graphics}
\usepackage{xcolor}
\usepackage{dirtytalk}
\usepackage[british]{babel}
\usepackage[utf8,latin1]{inputenc}
\UseRawInputEncoding
\usepackage{csquotes}
\usepackage{comment}
\usepackage[export]{adjustbox}
\usepackage{siunitx}
\usepackage[version=4]{mhchem}
\usepackage{amsmath}
\usepackage{amssymb}
\usepackage{mathtools}

\hypersetup{colorlinks,
  linkcolor=blue,%
  citecolor=blue,%
  urlcolor=blue}

\begin{document}

\title{Acoustic phonon-restricted four-phonon interactions: Impact on thermal and thermoelectric transport in monolayer \ce{h-NbN}}

\author{Himanshu Murari}
\affiliation{Department of Physics, Indian Institute of Technology Guwahati, Guwahati 781039, Assam, India}
\author{Subhradip Ghosh}
\affiliation{Department of Physics, Indian Institute of Technology Guwahati, Guwahati 781039, Assam, India}
\author{Mukul Kabir}
\affiliation{Department of Physics, Indian Institute of Science Education and Research, Pune 411008, India}
\author{Ashis Kundu}
\email[]{ashiskundu174@gmail.com}
\affiliation{Department of Physics, Indian Institute of Science Education and Research, Pune 411008, India}

\date{\today}

\begin{abstract}  
To explore the thermal and thermoelectric potential of 2D materials, we study the \ce{h-NbN} monolayer, which lacks mirror symmetry and features a large acoustic-optical phonon gap and quadratic flexural mode. First-principles calculations and the Boltzmann transport formalism reveal a complex interplay of multi-phonon scattering processes, where flexural phonons and four-phonon interactions play a significant role in heat transport, primarily dominated by acoustic phonons. Notably, the four-phonon interactions are predominantly confined to acoustic phonons. Tensile strain preserves the underlying scattering mechanisms while reducing anharmonicity and, consequently, the scattering rates, enhancing thermal conduction. Simultaneously, competing modifications in thermal and electrical transport shape the strain-dependent thermoelectric response, achieving a figure of merit $\sim$0.70 at elevated temperatures, a testament to its thermoelectric promise. Our findings underscore the critical role of microscopic transport modeling in accurately capturing thermal and thermoelectric properties, paving the way for advanced applications of 2D materials.
\end{abstract}


\maketitle

\section{Introduction\label{intro}}
Fundamental understanding of low-dimensional physical phenomena is critical for advancing device miniaturization, where quantum confinement significantly alters the electronic density of states and vibrational properties, thereby impacting transport behavior~\cite{dresselhaus2007new,zhang2017manipulation}. Thermal conductivity $\kappa$, influenced by enhanced phonon scattering in low-dimensional systems~\cite{Fugallo2015Phonon}, is pivotal for nanoelectronics. High $\kappa$ is essential for heat dissipation in power electronics, while low $\kappa$ optimizes thermoelectric energy conversion~\cite{goldsmid2013thermoelectric,ziman2001electrons}. Motivated by these considerations, research has expanded beyond graphene~\cite{graphene}, despite its high mobility, exceptional $\kappa$, and quantum Hall effect, due to its semimetallic nature that limits performance in electronic applications~\cite{gr_mobility,grph_kl,geisenhof2021quantum}. This has driven interest in alternative 2D semiconductors such as hexagonal boron nitride (h-BN)~\cite{hBN}, transition metal dichalcogenides~\cite{mos2_synth}, group-VA elements~\cite{li2014black}, and IV-VI monochalcogenides~\cite{zhao2014ultralow}, offering tunable band gaps suitable for ferroelectrics~\cite{fei2018ferroelectric}, thermoelectrics~\cite{zhao2014ultralow}, spintronics~\cite{spintronic}, and optoelectronics~\cite{YangLi,radisavljevic2011single}. In all such systems, tuning $\kappa$ is a crucial design strategy, as it governs their functional applicability in next-generation nanodevices.

High phonon scattering and low $\kappa$ in many 2D materials arise from their planar structure~\citep{nmat3064,RevModPhys.90.041002}. In particular, the out-of-plane flexural acoustic (ZA) modes, which exhibit quadratic dispersion, experience enhanced scattering~\citep{flexural_graph}.
Strong anharmonicity leads to frequent phonon-phonon interactions, significantly reducing phonon lifetimes~\citep{PhysRevB.87.214303}. Additionally, the high surface-to-volume ratio increases scattering from defects, impurities, and edges~\citep{PhysRevB.78.205403}. At elevated temperatures, Umklapp processes become increasingly active, further limiting thermal transport~\citep{KLEMENS19581}. Collectively, these mechanisms result in significantly suppressed $\kappa$ compared to bulk counterparts.

Modeling $\kappa$ using only three-phonon scattering processes is inadequate for 2D materials, especially when an acoustic-optical (A-O) phonon gap exists, flexural (ZA) modes are prominent, selection rules are strict, and acoustic phonon bunching arises from simplified lattice structures~\citep{RevModPhys.90.041002}. These features significantly enlarge the phase space for four-phonon interactions, opening additional scattering channels and leading to a substantial reduction in $\kappa$ for materials such as graphene~\citep{feng,PhysRevB.108.L121412}, MoS$_2$~\citep{PhysRevB.104.115403,gp_das}, and other 2D systems~\citep{PhysRevB.108.214304,4ph_grp_iv}. In graphene, theoretical predictions of high $\kappa$ exceeding \SI{3000}{\watt\per\meter\per\kelvin} at \SI{300}{\kelvin} arise from the long mean free paths of low-frequency ZA phonons, which dominate heat transport~\citep{grph_kl,flexural_graph,feng}. 
This is due to strict selection rules that limit three-phonon processes, especially for ZA modes, where in-plane reflection symmetry forbids interactions such as ZA + ZA $\leftrightarrow$ ZA~\citep{flexural_graph}.  However, inclusion of four-phonon scattering, driven by fourth-order anharmonicity, opens additional relaxation channels and significantly increases scattering rates, reducing $\kappa$ closer to experimental measurements of 1000--1500 \si{\watt\per\meter\per\kelvin}~\citep{feng,PhysRevB.108.L121412,10.1063/1.4796177,ncomms4689}. It is important to note that heat conduction in graphene is highly sensitive to structural defects, sample size, substrate interactions, and isotope composition~\citep{nl0731872,ncomms14486,C6NR03470E,ncomms4689,nl9041966,nmat3207} and is often associated with significant experimental uncertainties~\citep{nl9041966,nmat3207,nn102915x}. Each of these factors can lead to additional phonon scattering channels, thereby reducing the lattice thermal conductivity compared to the pristine case.

In the quest of novel 2D materials for next-generation devices, polymorphs of \ce{NbN}, the rectangular \ce{s-NbN} and honeycomb \ce{h-NbN} monolayers, have gained significant attention~\cite{Anand_NL16}. While \ce{s-NbN} retains superconducting behavior akin to its bulk rocksalt counterpart~\cite{Hazra_2016}, \ce{h-NbN} exhibits semiconducting properties along with an unusually large piezoelectric response, forbidden by symmetry in the bulk. Remarkably, \ce{h-NbN} also hosts a Dirac semimetallic state that coexists with ferroelectricity, which couples strongly to both strain and external electric fields~\cite{Chanana_PRL19}. Owing to broken inversion symmetry and preserved time-reversal symmetry, \ce{h-NbN} has been proposed as a promising valleytronic material~\cite{Ahammed_PRB22}. The interplay between valley and spin degrees of freedom, mediated by Zeeman and Rashba-type spin splittings, further enhances its potential for device applications. Despite extensive interest in its semiconducting behavior, the lattice thermal and electronic transport properties of \ce{h-NbN} remain largely unexplored.

We present a comprehensive investigation of the lattice thermal conductivity and electronic transport properties of the recently discovered honeycomb \ce{h-NbN} using first-principles density functional theory combined with phonon Boltzmann transport formalism. Our study reveals that four-phonon scattering plays a critical role in suppressing $\kappa_{l}$, even in the absence of mirror symmetry. Under mechanical strain, $\kappa_{l}$ shows a slight increase, primarily attributed to the changes in anharmonicity. Notably, $\kappa_{l}$ of \ce{h-NbN} remains significantly lower than that of prototypical 2D materials such as \ce{MoS2}. We further evaluate the electronic transport coefficients and thermoelectric figure of merit ($zT$) using an electron-phonon interaction-based approach. Our analysis reveals that four-phonon scattering plays a critical role in shaping $zT$, highlighting the necessity of its inclusion in predictive transport modeling. These findings indicate that \ce{h-NbN} is a promising candidate for thermoelectric and nanoelectronic applications, where multi-phonon processes govern overall performance.

\section{Computational details}

We employ density functional theory (DFT) to compute harmonic and anharmonic interatomic force constants, which are then used in the phonon Boltzmann transport equation (BTE) to evaluate $\kappa_l$, accounting for both three-phonon and four-phonon scattering processes, which are crucial for accurate estimation. The thermoelectric figure of merit $zT$ is determined by incorporating electron-phonon interactions into the electronic BTE to compute electrical conductivity $\sigma$, Seebeck coefficient $S$, and electronic thermal conductivity $\kappa_e$, offering a comprehensive evaluation of thermoelectric performance.

\subsection{Electronic structure}

First-principles calculations were performed using DFT~\citep{hohenberg1964inhomogeneous,kohn1965self} within the projector augmented wave (PAW) framework, as implemented in the Vienna Ab initio Simulation Package (VASP)~\citep{kresse1996efficient}. The exchange-correlation interactions were treated using the Perdew-Burke-Ernzerhof (PBE) functional within the generalized gradient approximation (GGA)~\cite{perdew1996generalized}. A plane-wave energy cutoff of 550 eV was employed, and the Brillouin zone was sampled using a 16$\times$16$\times$1 Monkhorst-Pack $k$-point grid~\cite{monkhorst1976special}. Structures were optimized until the total energy and atomic forces converged below 10$^{-7}$ eV and 10$^{-3}$ eV/\AA, respectively. A vacuum spacing of 20 \AA\ was introduced along the out-of-plane $c$-direction to eliminate interlayer interactions.

\subsection{Phonon transport}
Using the linearized phonon BTE, the lattice thermal conductivity is evaluated as,
\begin{equation}
    \kappa^{\alpha\beta}_{l} = \frac{1}{V}\sum_{\lambda} C_{\lambda} v^{\alpha}_{\lambda} v^{\beta}_{\lambda} \tau_{\lambda},
\end{equation}
where $V$ is the unit cell volume, $\alpha$ and $\beta$ denote Cartesian directions, $\lambda$ indexes the phonon modes, $C_{\lambda}$ is the mode-specific heat capacity, $v^{\alpha}_{\lambda}$ is the phonon group velocity along the $\alpha$ direction, and $\tau_{\lambda}$ is the phonon relaxation time. The phonon relaxation times were evaluated within the relaxation-time approximation (RTA) and then used in the iterative scheme, as implemented in the FourPhonon package~\cite{fourph}. However, the four-phonon relaxation times were included only at the RTA level in the iterative scheme.
The total scattering rate $\tau^{-1}_{\lambda}$ is determined using  Matthiessen's rule,
\begin{equation}
\frac{1}{\tau_{\lambda}} = \frac{1}{\tau_{{\rm 3 ph},\lambda}} + \frac{1}{\tau_{{\rm 4 ph}, \lambda}} + \frac{1}{\tau_{\rm iso}},
\label{eq2}
\end{equation}
where $\tau^{-1}_{{\rm 3ph},\lambda}$, $\tau^{-1}_{{\rm 4ph},\lambda}$ and $\tau^{-1}_{\rm iso}$~\cite{Tamura_PRB83} account for the three-phonon, four-phonon and phonon-isotopic scattering, respectively. Three-phonon and four-phonon scattering terms are given by~\cite{li2014shengbte,feng,fourph},
\begin{equation}
\scalebox{0.90}[1]{$
\tau_{{\rm 3ph}, \lambda}^{-1} = \displaystyle\sum_{\lambda_1, \lambda_2} \left[\frac{1}{2} \left(1+n_{\lambda_1}^0+n_{\lambda_2}^0\right) \Gamma_{-} + \left(n_{\lambda_1}^0 - n_{\lambda_2}^0\right) \Gamma_{+}\right]
$}
\label{3ph}
\end{equation}
\begin{eqnarray}
\scalebox{0.90}[1]{$
\begin{split}
\tau_{{\rm 4ph}, \lambda}^{-1}  = \displaystyle\sum_{\lambda_1, \lambda_2, \lambda_3} & \left[ \frac{1}{6} \frac{n_{\lambda_1}^0 n_{\lambda_2}^0 n_{\lambda_3}^0}{n_{\lambda}^0} \Gamma_{--} + \frac{1}{2} \frac{\left(1 + n_{\lambda_1}^0\right) n_{\lambda_2}^0 n_{\lambda_3}^0}{n_{\lambda}^0} \Gamma_{+-} \right. \\
& \left.  + \frac{1}{2} \frac{\left(1 + n_{\lambda_1}^0\right)\left(1 + n_{\lambda_2}^0\right) n_{\lambda_3}^0}{n_{\lambda}^0} \Gamma_{++} \right],
\end{split}
$}
\label{4ph}
\end{eqnarray}
where $n^{0}_{\lambda}$ is the Bose-Einstein distribution function depends on the phonon frequency. The terms $\Gamma_{\pm}$ and $\Gamma_{\pm\pm}$ represent the scattering probability matrices for three-phonon and four-phonon processes, respectively. These probabilities are evaluated using Fermi's golden rule and are constrained by both energy and momentum conservation.

The harmonic second-order and anharmonic third- and fourth-order interatomic force constants (IFCs) were computed using the finite-displacement method within DFT. Second-order IFCs were obtained using a 10$\times$10$\times$1 supercell and a 2$\times$2$\times$1 $k$-mesh using the Phonopy package~\cite{togo2015first}, with rotational sum rules enforced to correct symmetry-breaking artifacts~\cite{hiphive}. For third- and fourth-order IFCs, 8$\times$8$\times$1 supercells with 2$\times$2$\times$1 and $\Gamma$-only $k$-meshes were used, capturing interactions up to the 10$^{th}$ and 4$^{th}$ nearest neighbors, respectively. These IFCs were used to compute the lattice thermal conductivity $\kappa_{l}$ by iteratively solving the linearized PBTE using the FourPhonon code~\cite{fourph}. A dense $60 \times 60 \times 1$ $q$-mesh was employed to ensure convergence (see SI).

\subsection{Electronic transport and thermoelectric properties}
The thermoelectric figure of merit $zT$ is defined as, 
\begin{equation}
zT = \frac{S^{2}\sigma}{\kappa} T
\end{equation}
where $S$ is the Seebeck coefficient, $\sigma$ is the electrical conductivity, $T$ is the absolute temperature, and $\kappa = \kappa_e + \kappa_l$ is the total thermal conductivity, composed of electronic ($\kappa_e$) and lattice ($\kappa_l$) contributions~\citep{disalvo1999thermoelectric}. The electronic transport coefficients, $S$, $\sigma$, and $\kappa_{e}$, were evaluated by solving the BTE using the BoltzTrap2 package~\cite{madsen2018boltztrap2}. The energy-dependent carrier relaxation time ($\tau_{\rm ep}$) was determined using the electron-phonon averaged (EPA) approximation~\citep{epa}, which accounts for intrinsic electron-phonon interactions (see SI). EPA calculations were performed within the plane-wave pseudopotential framework of Quantum Espresso~\citep{QE-2009}, using ultrasoft pseudopotentials~\citep{uspp}. A kinetic energy cutoff of 100 Ry was employed, with $k$- and $q$-meshes of 16$\times$16$\times$1. Electronic eigenvalues were interpolated onto a finer 64$\times$64$\times$1 grid, and electron-phonon matrix elements were averaged over an energy grid with 0.6 eV spacing. Finally, the $zT$ values were obtained from the electronic and lattice transport contributions.

\section{Results and Discussion}
We begin with an analysis of the structural properties and phonon dispersion. We then examine the lattice thermal conductivity and its strain dependence, highlighting the essential role of four-phonon scattering in modeling heat transport. Finally, we explore the thermoelectric properties of the system.

\subsection{Structural details}

\begin{figure}[t]
\centering
\includegraphics[width=0.85\linewidth]{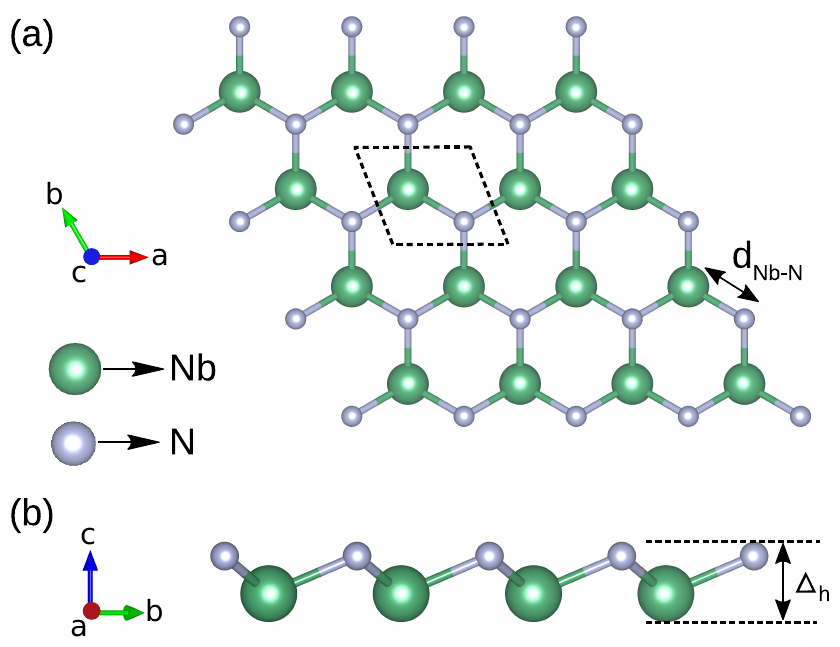}
\caption{
Crystal structure of \ce{h-NbN}. (a) Top view shows a hexagonal lattice. (b) Side view reveals significant buckling that breaks reflection symmetry, likely influencing heat transport. The short \ce{Nb-N} bond length indicates strong covalent bonding.
}
\label{fig1}
\end{figure}

The \ce{h-NbN} monolayer adopts a buckled hexagonal lattice (Figure~\ref{fig1}) derived from the [111] planes of bulk rocksalt \ce{NbN}. Unlike the planar honeycomb structures of graphene and \ce{h-BN}, \ce{h-NbN} features a pronounced buckling ($\Delta_{\rm h} = d_{\mathrm{Nb}}^z - d_{\mathrm{N}}^z$) of 0.77~\AA~ in the unstrained state, reflecting a vertical separation between \ce{Nb} and \ce{N} sublattices (Table~\ref{tab1}). This buckling breaks in-plane reflection symmetry, thereby altering phonon scattering and impacting $\kappa_l$. Similar symmetry-breaking effects have been reported in low-buckled 2D materials such as silicene, germanene, and stanene~\citep{4ph_grp_iv}. The optimized lattice constant of 3.15 \AA\ and \ce{Nb-N} bond length of 1.98 \AA\ indicate strong covalent bonding, consistent with earlier predictions~\citep{Chanana_PRL19, Ahammed_PRB22}. Heat conduction can be tuned by in-plane strain ($\epsilon_{xx} = \epsilon_{yy} = \epsilon$) as tensile strain reduces buckling (Table~\ref{tab1}). We investigate the effects of a tensile $\epsilon = 3\%$, remaining below the critical threshold at which the structure transitions to a planar configuration. The thermodynamic stability of monolayer h-NbN was verified through \textit{ab initio} molecular dynamics (AIMD) simulations at 300~K, 600~K, and 800~K. The structure remained intact with only minor distortions and no bond-breaking signs, confirming its stability at elevated temperatures (see details in SI).

The electronic structure of the \ce{h-NbN} monolayer reveals an indirect band gap of \SI{0.75}{\electronvolt} (\SI{0.78}{\electronvolt}) along the $K-\Gamma$ path using GGA-PBE (HSE06) functional (see SI), which agrees well with the previous HSE06 studies\cite{Ahammed_PRB22, NbX}. This moderate band gap confirms its semiconducting nature, suggesting that phonons dominate heat transport. Applying a tensile strain of $\epsilon = 3\%$ significantly narrows the band gap (Table~\ref{tab1}) and reduces buckling, which is expected to influence $\kappa_l$ by modifying phonon scattering mechanisms. Since the GGA-PBE and HSE06 functionals yield nearly identical band gaps and electronic features near the Fermi level, we have used the GGA-PBE functional throughout this work.


\begin{table}[t]
\caption{Structural properties of unstrained and strained \ce{h-NbN} monolayer. The lattice constant $a_{o}$, buckling height $\Delta_{\rm h}$, \ce{Nb-N} bond length $d_{\mathrm{Nb-N}}$, bond angle $\theta_{\mathrm{N-Nb-N}}$, and electronic band gap (E$_{g}$) are listed.}
\centering
\renewcommand{\arraystretch}{1.5}
\setlength{\tabcolsep}{5pt}
\resizebox{\linewidth}{!}{
\begin{tabular}{cccccc}
\hline
\hline
Strain $\epsilon$& $a_{0}$ (\AA) & $\Delta_{\rm h}$ (\AA) & $d_{\mathrm{Nb-N}}$ (\AA)& $\theta_{\mathrm{N-Nb-N}}$ ($^{\circ}$) & E$_{g}$ (eV) \\
    \hline 
    
       0\%  & 3.16  & 0.77   & 1.98  & 105.65 & 0.75\\
       3\%  & 3.25  & 0.70   & 2.00  & 108.46 & 0.40\\
\hline
\hline
\end{tabular}
}
\label{tab1}
\end{table}

\begin{figure*}[t]
    \centering
    \includegraphics[width=0.85\linewidth]{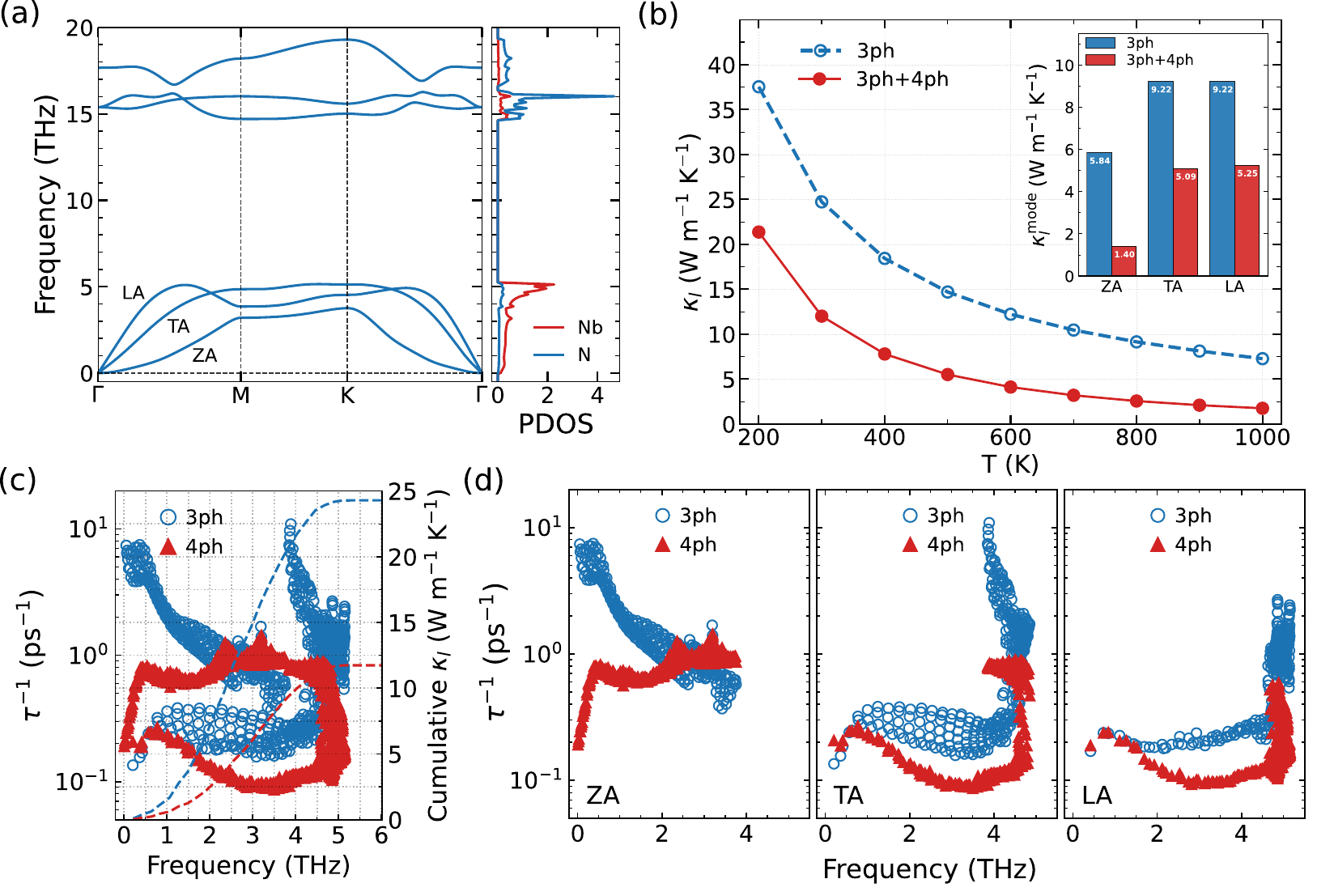}
    \caption{
(a) The phonon dispersion of \ce{h-NbN} confirms dynamic stability, featuring a quadratic ZA mode near $\Gamma$, a wide A-O gap, and nearly non-dispersive acoustic branches along $M-K$. PDOS analysis shows that acoustic modes are dominated by \ce{Nb} vibrations, while \ce{N} contributes mainly to optical modes. 
(b) The calculated lattice thermal conductivity ($\kappa_l$) shows a significant reduction when four-phonon scattering is included. The mode-resolved $\kappa_l^{\rm mode}$ at 300 K (inset) reveals that this suppression affects all acoustic branches, with the ZA mode being the most severely affected.
(c) Scattering rates involving acoustic modes calculated at 300 K indicate that four-phonon processes are weaker at low frequencies, but become comparable to three-phonon scatterings in the intermediate frequency range. Cumulative $\kappa_{l}$ corresponding to total scattering rates are also shown.
(d) Mode-resolved acoustic scattering rates show that the ZA mode predominantly governs phonon scattering. Although TA and LA modes exhibit stronger scattering at higher frequencies, their contributions to $\kappa_l$ remain limited.
}
\label{fig2}
\end{figure*}

\subsection{Lattice dynamics}
Phonon dispersion confirms the dynamic stability of \ce{h-NbN} monolayer, with no unstable modes [Figure~\ref{fig2}(a)]. A wide A-O phonon gap of \SI{9.5}{\tera\hertz} arises from the significant mass difference between \ce{Nb} and \ce{N}. The phonon density of states (PDOS) indicates that acoustic modes are associated with \ce{Nb} vibrations, while optical modes stem primarily from \ce{N} atoms. The flexural ZA mode exhibits the expected quadratic dispersion, $\omega \propto q^2$, near the $\Gamma$ point, while the in-plane transverse (TA) and longitudinal (LA) acoustic modes follow a linear $q$ dependence. Earlier, it was suggested that the quadratic ZA mode near $\Gamma$ in 2D materials is influenced by structural buckling, contrasting planar graphene with buckled silicene~\citep{non_quad}. However, it was later demonstrated that enforcing translational and rotational invariance in IFCs recovers the correct quadratic dispersion, independent of buckling~\citep{quadraticity}. By employing a larger supercell and enforcing the rotational sum rules, we ensure the quadratic ZA mode in \ce{h-NbN}, achieving improved accuracy over prior reports~\cite{Anand_NL16, Ahammed_PRB22}.

The quadratic ZA mode and wide A-O phonon gap in \ce{h-NbN} point to strong four-phonon scattering, consistent with observations in other 2D and bulk systems~\citep{feng, Zhang2023,BAs, Kundu_PRL21,snc}. This occurs because the large A-O gap suppresses many three-phonon scattering processes that cannot satisfy energy and momentum conservation. In contrast, four-phonon scattering processes are less restricted, as the involvement of an additional phonon offers greater flexibility in satisfying energy and momentum conservation laws. Moreover, the phonon dispersion of \ce{h-NbN} reveals nearly non-dispersive acoustic branches along the $M-K$ path, reflected in a pronounced PDOS peak near the top of the acoustic band [Figure~\ref{fig2}(a)]. Such flat dispersion further amplifies four-phonon scattering~\cite{non_dispersive}, reinforcing the need to account for higher-order anharmonicity in accurately evaluating the intrinsic lattice thermal conductivity.

\subsection{Impact of four-phonon scattering on $\kappa_l$}
Four-phonon interactions significantly reduce $\kappa_l$ across all temperatures [Figure~\ref{fig2}(b)]. At 300 K, the inclusion of both three- and four-phonon scatterings results in a 52\% decrease in comparison to that obtained using only three-phonon scattering. This reduction becomes more pronounced at higher temperatures, reaching 72\% at 800 K. This trend highlights the stronger temperature dependence of four-phonon scattering ($\tau^{-1}_{\rm 4ph} \sim T^2$) compared to the linear scaling of three-phonon scattering ($\tau^{-1}_{\rm 3ph} \sim T$). However, the impact of four-phonon processes varies across materials due to symmetry and phonon dispersion characteristics.  At \SI{300}{\kelvin},  the reduction in $\kappa_l$ is substantial in graphene and \ce{MoS2} ($\sim$75\%)~\citep{feng, gp_das}, moderate in stanene and \ce{SnC} ($\sim$50\%)~\citep{4ph_grp_iv, snc}, and modest in silicene and germanene ($\sim$35\%)~\citep{4ph_grp_iv}. In \ce{h-NbN}, acoustic phonons dominate heat transport, contributing about 98\% to $\kappa_l$ [Figure~\ref{fig2}(b)], consistent with other 2D materials. Mode-resolved $\kappa_l^{\rm mode}$ reveals that four-phonon scattering most strongly suppresses the ZA mode contribution [Figure~\ref{fig2}(b)], highlighting its critical role in heat transport. 

To further investigate the underlying mechanisms, we present the frequency-dependent scattering rates and cumulative $\kappa_l$ in the acoustic region  [Figure~\ref{fig2}(c)]. Notably, four-phonon scattering is significant and comparable to three-phonon scattering in the intermediate frequency ($2\text{-}4.5$ \si{\tera\hertz}) range. This trend persists at higher temperatures (see SI), where four-phonon scattering becomes dominant above \SI{1}{\tera\hertz}, consistent with its stronger temperature dependence. However, three-phonon scattering remains dominant at low frequencies across all temperatures. This behavior is in contrast with prototypical 2D materials such as graphene~\citep{feng}, \ce{MoS2}~\cite{gp_das}, and \ce{SnC}~\cite{snc}, where in-plane mirror symmetry enforces reflection symmetry selection rules (RSSR). These rules forbid three-phonon processes involving an odd number of ZA phonons (e.g., TA/LA + ZA $\leftrightarrow$ TA/LA and ZA + ZA $\leftrightarrow$ ZA), thus suppressing effective three-phonon scattering~\citep{flexural_graph,grp_kl,grph_kl}. In such cases, four-phonon processes become relevant, as they allow interactions involving two and four ZA phonons~\citep{feng}. Combined with the large phonon population arising from the quadratic ZA dispersion at low frequencies, this explains the dominance of four-phonon scattering in 2D materials with mirror symmetry. 

In buckled 2D materials, the absence of mirror symmetry lifts the constraints imposed by RSSR, allowing both three- and four-phonon processes involving any number of ZA modes. Within the framework of perturbation theory, higher-order interactions are typically weaker; however, a large A-O phonon gap, flexural modes, and nearly non-dispersive acoustic branches [Figure~\ref{fig2}(a)] together with enhanced four-phonon scattering in \ce{h-NbN} [Figures~\ref{fig2}(b) and \ref{fig2}(c)]. The A-O gap limits three-phonon processes involving optical modes due to restricted phase space for energy-momentum conservation, whereas four-phonon processes, with greater combinatorial flexibility, remain less restricted.

\begin{figure}[t]
\centering
\includegraphics[width=0.95\linewidth]{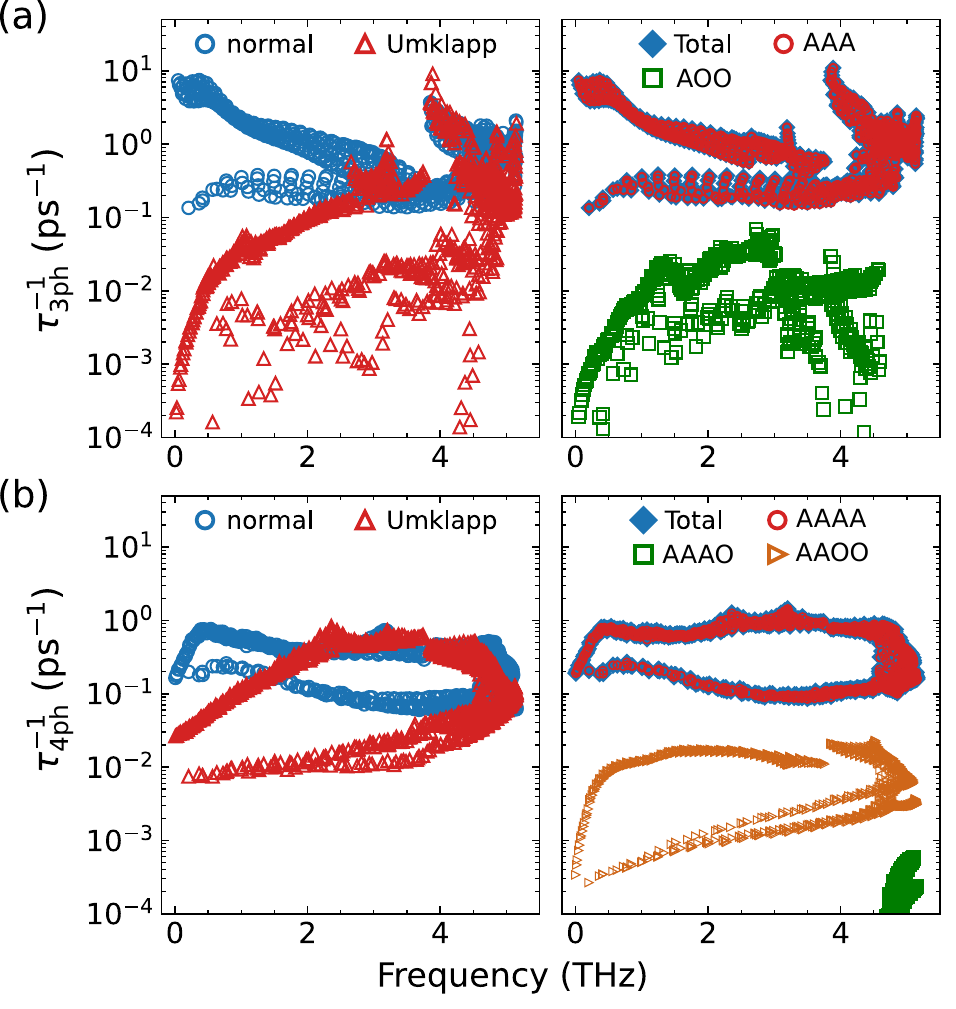}
\caption{(a) Three-phonon (3ph) (b) four-phonon (4ph) scattering rates are presented. At \SI{300}{\kelvin}, both three-phonon and four-phonon scattering in monolayer \ce{h-NbN} are predominantly governed by normal processes rather than Umklapp scattering.  Moreover, phonon scattering is primarily driven by all-acoustic processes, specifically the AAA and AAAA channels.
    }
\label{fig3}
\end{figure}

\begin{figure*}[t]
    \centering
    \includegraphics[width=0.65\linewidth]{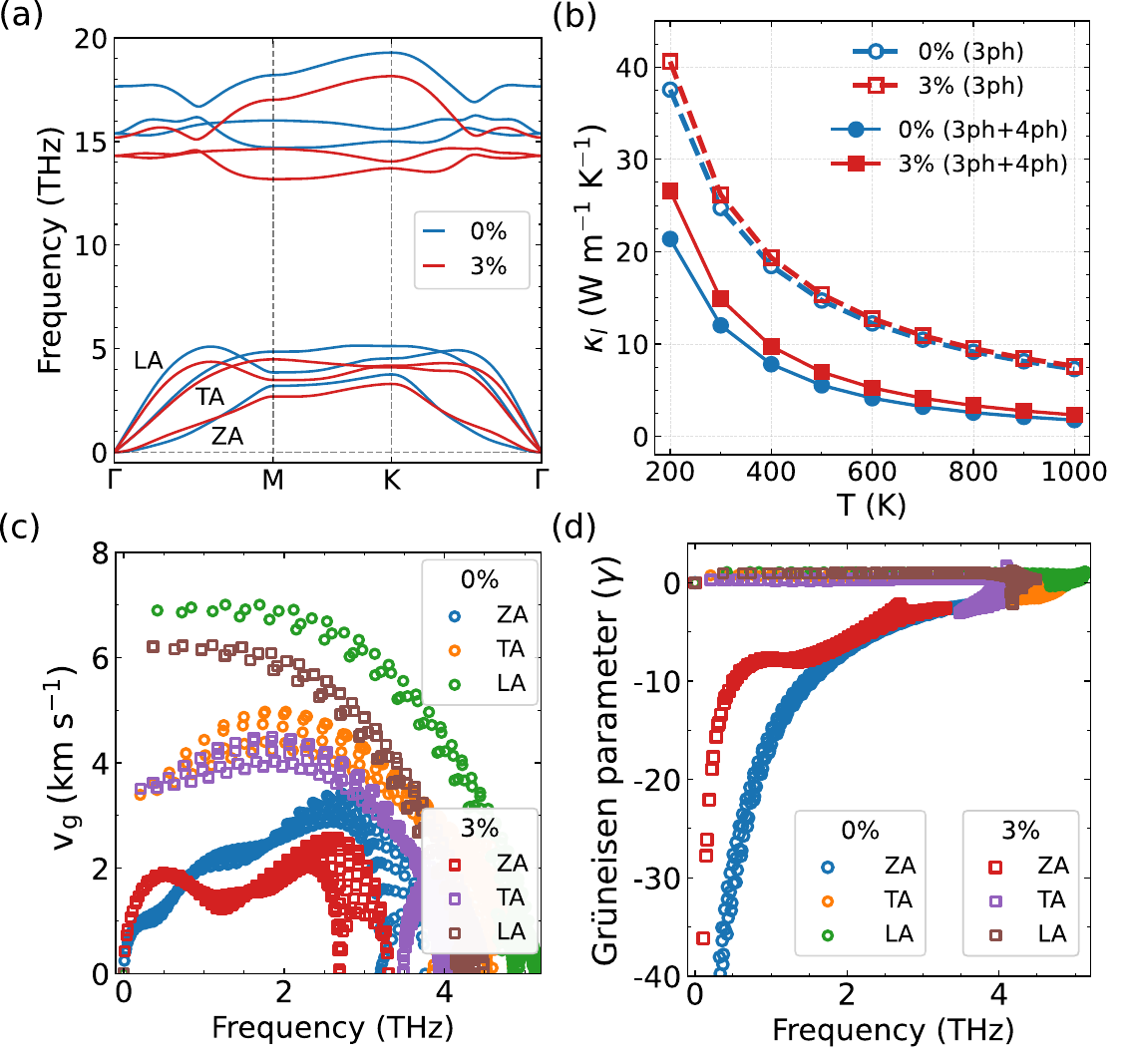}    
    \caption{    
(a) The \ce{h-NbN} monolayer remains dynamically stable under tensile strain $\epsilon = 3\%$, with phonon dispersion showing overall softening, a reduced A-O phonon gap, and a transition to nearly linear ZA dispersion near the Brillouin zone centre. 
(b) Lattice thermal conductivity ($\kappa_l$) increases under tensile strain across all temperatures. Similar to the unstrained case, four-phonon scattering remains equally significant under strain. 
(c) Phonon softening under strain reduces the group velocity $v_\text{g}$ of all acoustic modes, except for the ZA mode below \SI{1}{\tera\hertz}, due to its transition toward linear dispersion. 
Mode-resolved Gr\"uneisen parameter $\gamma$ reveals that anharmonicity is primarily governed by the ZA mode, whose contribution is significantly suppressed under tensile strain.}
\label{fig4}
\end{figure*}

Mode-resolved scattering rates provide further insight into thermal transport in \ce{h-NbN} [Figure~\ref{fig2}(d)]. Since $\kappa_l$ is dominated by acoustic modes, we focus on their $\tau^{-1}$. While low-frequency phonons generally exhibit long lifetimes due to limited phase space, ZA modes involved in three-phonon processes display anomalously short lifetimes, indicative of strong scattering. This arises from the high population of low-energy ZA phonons, owing to their quadratic dispersion and absence of mirror symmetry, which lifts RSSR constraints and permits all ZA involved processes. Consequently, the three-phonon scattering is dominated by normal processes [Figure~\ref{fig3}], which predominantly redistribute phonon momentum and contribute indirectly to thermal resistance by promoting subsequent Umklapp processes, which directly resist the heat flow by reversing phonon momentum~\cite{Maznev_AJP14}. A similar dominance of normal scattering has been reported in other low-buckled 2D materials like germanene and stanene~\citep{4ph_grp_iv}. However, in \ce{h-NbN}, despite strong three-phonon scattering involving the ZA mode, its contribution to $\kappa^{\rm 3ph}_l$ remains substantial, albeit lower than those of the TA and LA modes. This contrasts with other 2D materials~\cite{gp_das,feng}, where the ZA mode is the dominant contributor to $\kappa_l$ when only three-phonon scattering is considered.

Four-phonon scattering of the ZA mode remain stronger than that of the TA and LA modes [Figure~\ref{fig2}(d)], mainly via normal scattering [Figure~\ref{fig3}(b)], reducing $\kappa_l^{\rm ZA}$ nearly fourfold, while the contributions from the TA and LA modes experience a smaller suppression. This trend in $\kappa_l^{\rm mode}$ now resembles that of graphene and other 2D materials~\citep{feng}. Another key feature is the strong $\tau^{-1}_{\rm TA}$ and $\tau^{-1}_{\rm LA}$ in the $4\text{-}5$ \si{\tera\hertz} range [Figure~\ref{fig2}(d)], driven by a high PDOS from non-dispersive acoustic branches in this region [Figure~\ref{fig2}(a)]. However, this has minimal impact on $\kappa_l$, as the cumulative $\kappa_l$ reveals that low-frequency acoustic phonons primarily govern thermal conductivity [Figure~\ref{fig2}(c)].

The prominence of all-acoustic (AAA and AAAA) scattering in multi-phonon processes (Figure~\ref{fig3}) arises from the quadratic ZA mode, large A-O phonon gap, and nearly flat acoustic branches. This behavior contrasts with bulk materials~\cite{BAs, Kundu_PRL21}, where mixed acoustic-optical AAOO interactions typically dominate, but aligns with graphene lacking a phonon gap~\cite{feng} and \ce{AgCrSe2} featuring flat dispersions~\citep{non_dispersive}.

\subsection{Effect of tensile strain on $\kappa_{l}$}
A practical route to modulate ZA mode is through tensile strain, which can linearise its otherwise quadratic dispersion even at small values~\cite{grph_kl}. This transition significantly alters phonon populations and scattering dynamics, thereby impacting heat transport.  To explore this effect, we apply a modest in-plane tensile strain of 3\%, well below the threshold for inducing a planar structural transition~\cite{Chanana_PRL19}.

The strained structure remains dynamically stable, with all phonon branches softening and shifting to lower frequencies [Figure~\ref{fig4}(a)], leading to modified group velocities. Notably, the ZA mode transitions from a quadratic to a near-linear dispersion near the $\Gamma$ point, consistent with observations in other 2D materials~\citep{snc,gp_das}. Strain induces stronger softening of the optical branches, slightly narrowing the A-O gap (\SI{8.70}{THz}), which is unlikely to influence $\kappa_l$. Additionally, it slightly enhances acoustic phonon bunching, which can modify scattering rates.

Regardless of scattering order, the strained lattice exhibits a higher $\kappa_l$, with four-phonon processes being most significantly impacted [Figure~\ref{fig4}(b)]. At \SI{300}{\kelvin}, $\kappa_l$ rises by 23\% to \SI{14.80}{\watt\per\meter\per\kelvin}, when accounting for scattering up to the fourth order; a notable increase compared to the result obtained by considering only three-phonon interactions. Mode-resolved analysis (see SI) reveals that acoustic modes remain the primary contributors to both $\kappa^{\rm 3ph}_l$ and $\kappa^{\rm 3ph+4ph}_l$, with similar trends observed in the unstrained case.

\begin{figure*}[t]
\centering
\includegraphics[width=0.95\linewidth]{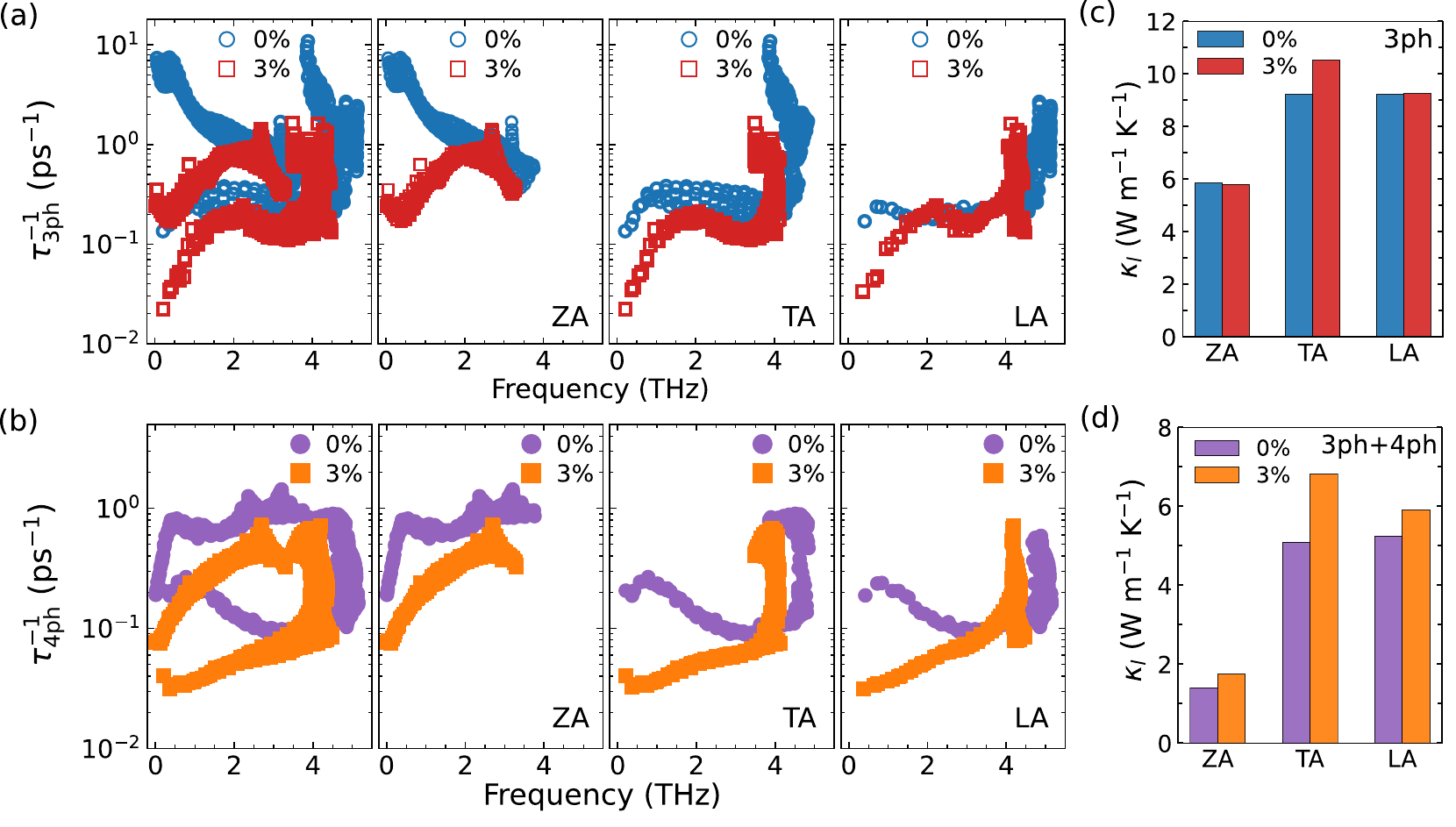}    
\caption{
(a) Three-phonon $\tau^{-1}_{\rm 3ph}$ and (b) four-phonon $\tau^{-1}_{\rm 4ph}$ scattering rates and their mode-resolved contributions, in the strained monolayer, exhibit a pronounced reduction in the low frequency range, which plays a dominant role in heat conduction. Mode-resolved $\kappa_l$ for (c) ${\rm 3ph}$ and (d) ${\rm 3ph} + {\rm 4ph}$ processes shows that the modest enhancement is primarily driven by the weakening of four-phonon scattering under strain. All results presented are computed at \SI{300}{\kelvin}.}
\label{fig5}
\end{figure*}

We analyze the group velocity $v_\text{g}$, Gr\"uneisen parameter, and phonon scattering rates to elucidate the mechanism driving thermal transport under strain. Tensile strain generally lowers $v_\text{g}$ ($= d\omega/dq$) across all frequencies [Figure~\ref{fig4}(c)], consistent with phonon softening and a narrower acoustic bandwidth [Figure~\ref{fig4}(a)]. An exception arises near the Brillouin zone centre, where the ZA mode exhibits an increase in $v_\text{g}$ below \SI{1}{\tera\hertz} as the dispersion becomes nearly linear, a regime that contributes negligibly to $\kappa_l$. Given that $\kappa_l \propto C v_\text{g}^2 \tau$ (with $C$ as heat capacity), we compute the small-grain $\kappa_l$ to isolate the influence of $v_\text{g}$, which drops from 2.2 to 1.8 \si{\watt\per\meter\per\kelvin\per\nano\meter} under strain, consistent with reduced $v_\text{g}$. However, the overall increase in $\kappa_l$ suggests that anharmonic phonon scattering plays the dominant role. To probe this, we calculate the Gr\"uneisen parameter $\gamma$, a measure of anharmonicity. Mode-resolved $\gamma$ reveals that ZA phonons exhibit the highest anharmonicity in unstrained \ce{h-NbN} [Figure~\ref{fig4}(d)], aligning with their strong scattering. Under strain, $\gamma_{_{\rm ZA}}$ reduces by a factor of 2.2, indicating suppressed anharmonicity, and thus lower scattering rates, ultimately enhancing $\kappa_l$. Although the decrease in $\gamma$ is less pronounced for LA and TA modes, they follow the same decreasing trend (see SI). This trend aligns with graphene and \ce{SnC} but differs from \ce{MoS2}~\cite{grph_kl,snc,gp_das}.

\begin{figure*}[t]
\centering
\includegraphics[width=0.95\linewidth]{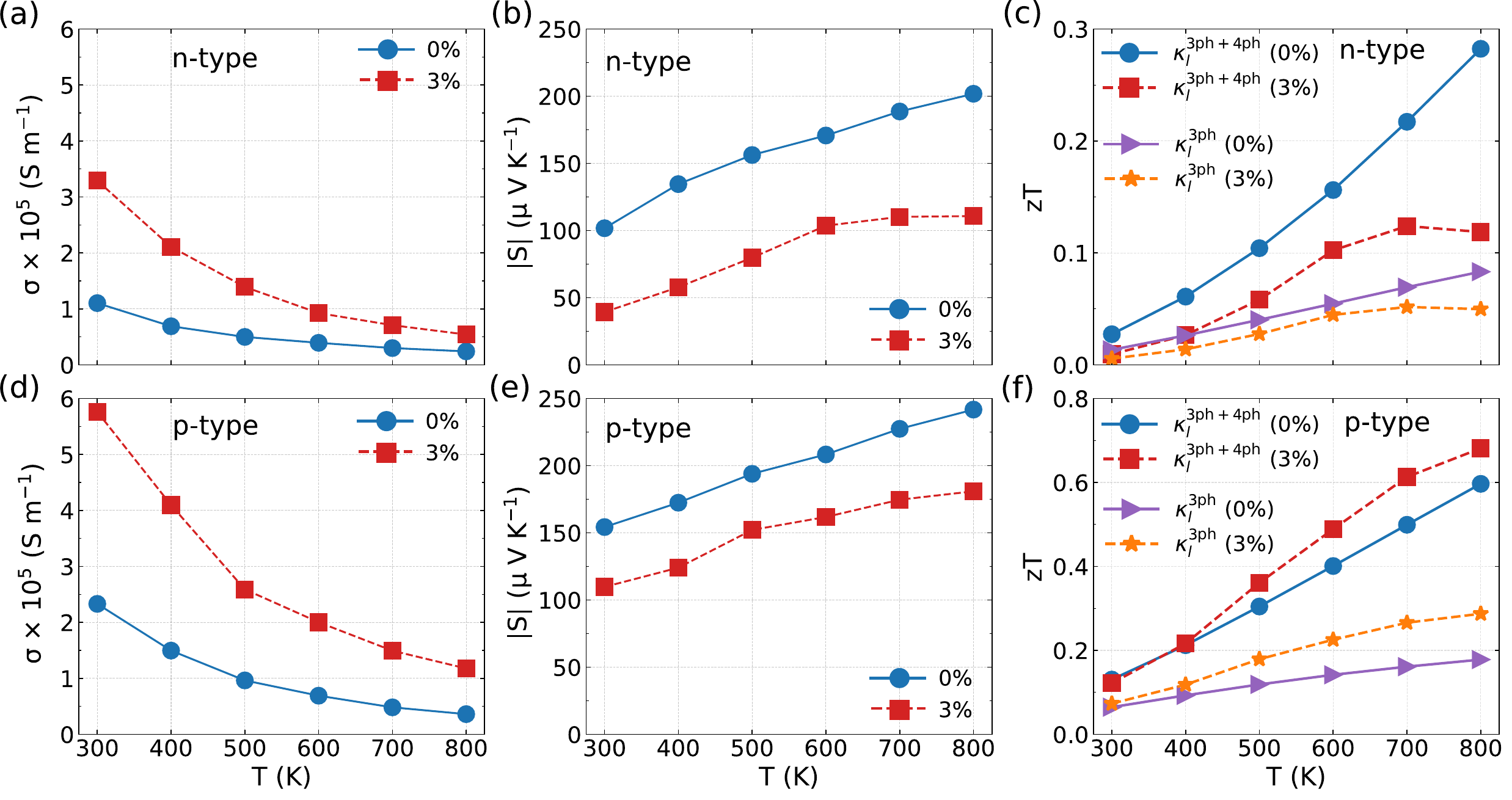}    
\caption{
Electrical conductivity $\sigma$ improves under tensile strain for both (a) n-type and (d) p-type carrier doping through a reduction in the electronic band gap. (b) and (e) As expected, the Seebeck coefficient $S$ exhibits an inverse trend to $\sigma$ under tensile strain. (c) and (f) The thermoelectric figure of merit $zT$, evaluated at a carrier concentration of $\sim 10^{20}$ cm$^{-3}$, exhibits distinct strain-dependent behavior for n-type and p-type doping. Calculations using both $\kappa^{\rm 3ph}_{l}$ and $\kappa^{\rm 3ph+4ph}_{l}$ highlight the critical role of four-order phonon scattering in accurately predicting $zT$.}
\label{fig6}
\end{figure*}

To further elucidate the scattering mechanism, we compute mode-resolved $\tau^{-1}$ for the strained lattice and compare it with the unstrained monolayer [Figure~\ref{fig5}(a) and~\ref{fig5}(b)]. Across all acoustic modes, tensile strain consistently reduces $\tau^{-1}$, regardless of the scattering order, in line with the previously discussed reduction in anharmonicity. This decrease in scattering rates accounts for the enhanced $\kappa_l$ observed in strained \ce{h-NbN} [Figure~\ref{fig4}(b)]. 

Mode-specific contributions to $\kappa_{l}$ offer valuable insight. Although the low-frequency scattering in $\tau^{-1}_{\rm 3ph}$ and $\tau^{-1}_{\rm 4ph}$ is significantly reduced under strain, the regime below $\sim$ 2 THz is primarily governed by normal processes. In the 2-5 THz range, however, the contributions of Umklapp and normal processes are comparable. The combined effect of the increased group velocity of the ZA mode (below $\sim$ 2 THz) [Figure~\ref{fig4}(c)] and the reduced scattering rates [Figure~\ref{fig5}(a)] in this same frequency range results in only a marginal change in its contribution to $\kappa_{l}$. Consequently, $\kappa_{l}$ under three-phonon scattering remains largely unaffected [Figure~\ref{fig5}(c)]. An exception is the TA mode, where a reduction in $\tau^{-1}_{\rm 3ph}$ extends into the intermediate frequency range, leading to a modest increase in $\kappa_{l}$ [Figure~\ref{fig4}(b)]. In contrast, a pronounced reduction in $\tau^{-1}_{\rm 4ph}$ across the intermediate frequency range, which governs heat transport, leads to an overall increase in mode-resolved $\kappa_{l}$ for all acoustic modes [Figure~\ref{fig5}(d)].

Similar to the unstrained monolayer (Figure~\ref{fig3}), the multi-phonon scattering is predominantly driven by all acoustic modes (see SI). For three-phonon scattering, decay ($\lambda_{1} \rightarrow \lambda_{2} + \lambda_{3}$) and absorption ($\lambda_{1} + \lambda_{2} \rightarrow \lambda_{3}$) processes contribute comparably in both unstrained and strained \ce{h-NbN} (see SI). In contrast, the four-phonon redistribution process ($\lambda_{1}+\lambda_{2} \leftrightarrow  \lambda_{3}+\lambda_{4}$) consistently overweights splitting ($\lambda_{1}\rightarrow \lambda_{2}+\lambda_{3}+\lambda_{4}$) and recombination ($\lambda_{1}+\lambda_{2}+\lambda_{3}\rightarrow \lambda_{4}$), irrespective of strain (see SI). The dominance of the redistribution process and AAAA scattering highlights ZA $+$ ZA $\leftrightarrow$ ZA $+$ ZA as the leading scattering channel. The reduction in scattering rates under tensile strain arises from the contraction of the three- and four-phonon phase space (see SI), leading to a modest increase in $\kappa_l$, albeit less pronounced than in other 2D materials~\citep{snc}. These findings highlight the critical role of higher-order phonon interactions in accurately modeling thermal transport. 
Compared to other monolayers, the $\kappa_l$ of \ce{h-NbN} is lower than that of materials such as \ce{MoS2}~\cite{gp_das}, Janus monolayers like \ce{PtSTe}~\cite{Pan_PRM22} and \ce{WSSe}~\cite{Patel_ACSAMI20}, and \ce{TiS3}~\cite{Zhang_ACSAMI17}, yet remains higher than that of low-$\kappa_l$ 2D energy materials such as \ce{SnSe}~\cite{Jin_CASEL18} and \ce{SnS}~\cite{Aseginolaza_PRB19}.

\subsection{Thermoelectric properties}

Owing to the intrinsically low lattice thermal conductivity of \ce{h-NbN}, it is pertinent to evaluate its thermoelectric properties. We evaluate key parameters, including the electrical conductivity $\sigma$, Seebeck coefficient, and figure of merit $zT$ (Figure~\ref{fig6}), with particular emphasis on the influence of four-phonon scattering and strain. Incorporating energy-dependent electron-phonon interactions to calculate carrier relaxation times (see SI), even when averaged, is crucial for accurately computing electronic transport properties. This approach, combined with a comprehensive thermal transport modeling framework, outperforms conventional deformation potential methods, which often overestimate $zT$ by oversimplifying relaxation times.

The narrowing of the electronic band gap under tensile strain (Table~\ref{tab1}) results in an increase in $\sigma$, accompanied by a corresponding decrease in the Seebeck coefficient $S$ (Figure~\ref{fig6}). Although both carrier types follow a similar trend, the p-type system exhibits significantly higher $\sigma$, attributed to the more dispersive valence band (see SI), yielding a lighter effective mass near the band edge.  

Accounting for four-phonon scattering is crucial, as it reduces $\kappa_l$, and prevents underestimation of $zT$ if neglected
[Figure~\ref{fig6}(c) and (f)]. For n-type doping, $zT$ decreases under tensile strain due to the rising $\kappa_l$. In contrast, p-type doping significantly enhances $zT$, even altering its strain dependence, as the enhancement in $S^2\sigma$ outweighs the increase in $\kappa_l$. Given that the optimal $zT$ is determined at a carrier concentration near the Fermi level, high $zT$ values can be attained through moderate carrier doping. At higher temperatures, the maximum calculated $zT$ of \ce{h-NbN} is $\sim$0.7, comparable to that of Janus monolayers of PtSTe ($\sim$0.6)~\cite{Pan_PRM22} and WSTe ($\sim$0.74)~\cite{Patel_ACSAMI20}, highlighting its potential for high-temperature thermoelectric applications.

\section{Conclusions}

We present a comprehensive first-principles study of phonon and electronic transport in two-dimensional \ce{h-NbN}, combining density functional theory with Boltzmann transport formalism. Our findings highlight the crucial role of multi-phonon scattering and strain in governing thermal and thermoelectric performance. Even in the absence of in-plane reflection symmetry selection rules due to lattice buckling, four-phonon scattering significantly limits lattice thermal conductivity. This arises predominantly from all-acoustic AAAA processes due to a large acoustic-optical phonon gap and is further amplified by the quadratic ZA mode and weakly dispersive acoustic branches.

Although the underlying scattering mechanisms remain unchanged under tensile strain, reduced bonding covalency softens the acoustic phonons, leading to lower group velocities and Gr\"uneisen parameters. This diminished anharmonicity suppresses scattering rates, thereby enhancing the lattice thermal conductivity. Our results reveal that the thermoelectric response under tensile strain is governed by concurrent increases in thermal and electrical conductivity, driven by electronic band gap reduction. A high-temperature $zT$ of $\sim$0.7 highlights the promising thermoelectric potential of \ce{h-NbN}. These findings underscore the importance of comprehensive microscopic modeling, not only to accurately capture thermal transport, but also to reliably predict thermoelectric performance.



\section*{Acknowledgements}
A. K. acknowledges support from the CHANAKYA Post-Doctoral Fellowship, funded by the National Mission on Interdisciplinary Cyber-Physical Systems (NM-ICPS), Department of Science and Technology (DST), Government of India, through the I-HUB Quantum Technology Foundation, Pune. A. K. and M. K. gratefully acknowledge the computational resources provided by the PARAM Brahma Facility at IISER Pune, under the National Supercomputing Mission of the Government of India. H. M. and S. G. thank DST for computational support under Grant No. SR/FST/P-II/020/2009, and acknowledge the PARAM supercomputing facilities at IIT Guwahati.

\sloppy


%

\end{document}